# Importance of strain reduction in improvement of optical transmission and conductance in Si$^{4+}$ doped ZnO: a probable new moisture resistant Transparent Conductive Oxide

**Tulika Srivastava[1], Aswin Sadanandan[1], Gaurav Bajpai[1], Saurabh Tiwari[1], Ruhul Amin[1], Mohd. Nasir[1], Sunil Kumar[1], Parasharam M. Shirage[1,2], Sajal Biring[3]\*\*, Somaditya Sen[1,2]\***

[1]Metallurgical and Material Sciences, Indian Institute of Technology Indore, Simrol Campus, Khandwa Road, Indore 452020 India

[2]Discipline of Physics, Indian Institute of Technology Indore, Simrol Campus, Khandwa Road, Indore 452020, India

[3]Electronic Engg., Ming Chi University of Technology, New Taipei City, Taiwan

**Email ids: \***sens@iiti.ac.in, **\*\***biring@mail.mcut.edu.tw

**Abstract:** Si doped ZnO has been reported to be a better conductor than pure ZnO. It is reported that carrier density increases and hence conductivity increases. However, the effect on optical transmission is yet not clear until our recent report [1]. $Zn_{1-x}Si_xO$ for x= 0, 0.013, 0.020 and 0.027 have been synthesized using sol-gel method (a citric acid-glycerol route) followed by solid state sintering. We found that there is a decrease in defect states due to Si doping. The correlation of strain to the decrement in vacancy sites is discussed in this report. In modern electronics and solar cell fabrication, transparent conductive oxides (TCOs) are important components which conduct electrically without absorbing visible light. Known TCOs are extremely costly and are composed of non-abundant elements. Search for new ecofriendly, cheap and sustainable TCOs has been a recent research of attraction. Keeping in mind that most solar cells are exposed to natural conditions, the humidity tolerance becomes a determining factor. We report that sensitivity to moisture decreases drastically while the conductivity and optical transparency increases with doping. The reduction of strain and improvement of transport properties results in increased conductivity of Si doped ZnO pellets, tempting us to envisage this material as a probable alternate TCO.

**Keywords:** Si doped ZnO, transparent conductive oxide, humidity, conductivity, adsorption

## 1. Introduction:

ZnO is a very promising transparent conductive material to replace ITO and FTO because of its low cost, non-toxicity, good optical and electrical properties and high thermal stability. To further improve conductivity of ZnO, it is doped with group III, IV and V elements. Several reports are present on doped ZnO such as Hf [2], Ga [3], Al [4] etc. which shows improved electrical and optical properties and are fit for applications as a TCO. Residual Strain, which is produced by lattice mismatch and the thermal expansion coefficient differences [5], has a great impact on electronic and optical properties of materials. Hence, the strain distribution in the ZnO particles is an important subject to be investigated. Moreover, TCO materials should have stable electrical properties when exposed to humidity; i.e. they should be moisture resistant. When exposed to humidity, rapid degradation in electrical conductivity is observed in Al-doped ZnO [6, 7] and Ga-doped ZnO [8]. Miyata et al. [9] showed similar results reporting instability in high humidity condition.

Some reports suggested that silicon doped ZnO can be used as TCO due to its high optical transparency (~80-85% in visible region), high conductivity, high mobility and good thermal stability [10, 11].There are several solution based synthesis routes as reported [12-18] which may yield various morphologies depending upon precursors and synthesis conditions to explore more functionality. In this report, we prepare solgel processed samples by a citric acid route and analyze the effect of silicon doping on the strain and disorder in host ZnO lattice. Silicon doping reduces strain which improves its electrical & optical properties and makes it fit for TCOs application. However to claim this material as a robust TCO one needs to prove that the material is moisture resistant, i.e. its electrical properties does not get drastically changed under high humidity. Here, we have also investigated the effect of humid environment on the conductivity of $Zn_{1-x}Si_xO$ pellets.

## 2. Experimental:

$Zn_{(1-x)}Si_xO$ nanoparticles, ZS0, ZS1, ZS2 & ZS3 with x=0, 0.013, 0.020 & 0.027 respectively have been synthesized by sol-gel method (standard Pechini method) followed by solid state sintering. A precursor solution was prepared by dissolving ZnO powder in $HNO_3$ (Alfa Aesar, purity 99.9%). To this solution, appropriate amount of orthosilicate $[(C_2H_5O)_4Si]$ is added and stirred for some time for proper homogenous mixing. The gelling agent used in this process was citric acid and glycerol [19]. The two polymerize by releasing $H_2O$ from OH groups (esterification) of citric acid and glycerol when heated at 70°C for 1 hr. The polymeric solution was added to the Zn/Si solution. The resultant solution was stirred vigorously while being heated. Zn and Si ions get attached homogenously to the polymer solution. Evaporation of water in the solution resulted in a gel formation in about 4hrs. The gels were burnt on the hot plate in ambient conditions. The resultant powders were decarbonized and denitrified at 450°C for 6 hrs followed. Pellets were pressed at 4T pressure and sintered at 1000°C.

Structural characterization is carried out using x-ray diffractometer (Bruker D2-Phaser). Surface morphology is investigated by field emission scanning electron microscopy (Carl Zeiss FESEM Supra-55). Room temperature photoluminescence measurement was investigated using Dongwoo Optron DM 500i. Agilent UV-vis spectrometer (model Carry 60) was used to obtain transmission spectra and analyze electronic bandgaps. Vibrational modes were studied by Raman Spectroscopy (Research India Raman Spectroscope RIRM151). Conductivity measurement was done by Keithely source meter (model no. 2401). We performed humidity sensing using a home-made set up [Figure 1]. Two stainless steel electrodes connected to a Keithley Source Meter (Keithley 2401) were arranged with a separation of 1.25 mm, to measure the current-voltage (I-V) characteristics and humidity dependent resistance ($R_H$) of $Zn_{(1-x)}Si_xO$ pellets. The arrangement was enclosed inside a sealed glass chamber to create and control a humid atmosphere. A digital hygrometer (Maxtech TM-1) was placed inside this chamber to measure the relative humidity. The Relative Humidity (RH) under consideration in the "Off" and "On" modes are 45% to 75%. The sealed chamber had one inlet connected to a two-way key and one outlet connected to a vacuum pump. The two-way key had two inlets and one outlet. One inlet was for dry air and another for moisture saturated air. The outlet was connected to the inlet of the test chamber.

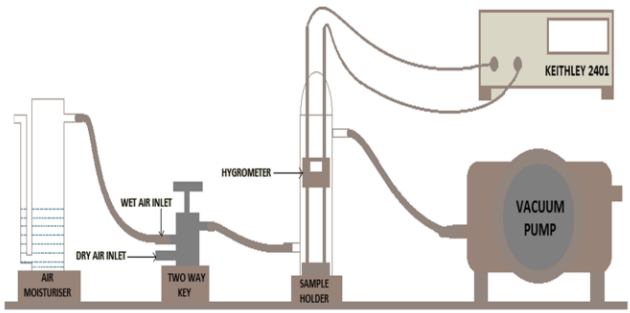

Figure 1: Humidity sensing set up

## 3. Results & Discussion:

SEM images (figure 2) reveal agglomerated spherical particles with average particle size increasing from 0.695-1.3 microns with silicon incorporation. Size was calculated using ImageJ software. It is well known that strain in the lattice decreases with increasing size of the particles. Hence one can expect reduction of lattice strain due to silicon incorporation.

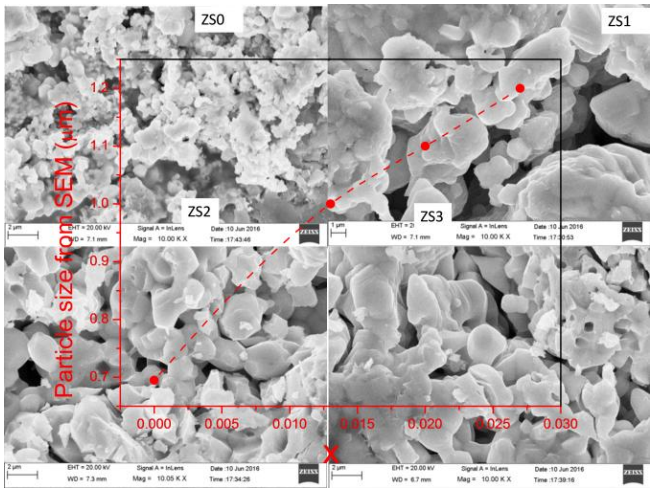

Figure 2: SEM image of ZS0, ZS1, ZS2 & ZS3 (inset is particle size calculation from imageJ software)

XRD spectra (figure 3(a)) reveal a dominant hexagonal wurtzite ZnO structure with some minor reflections similar to zinc blend. However, there is no secondary phase found related to simple or complex oxides of Zn and Si. We have calculated the average crystalline size of the particles from the XRD patterns using Debye Scherer's formula [20]. The crystalline size increases with substitution similar to the particle size from SEM from 55nm to 67nm as shown in Figure 3a(inset). This smaller size from XRD studies as compared to SEM studies is well interpreted as we can see agglomerated particles in SEM images being composed of smaller size crystals. The c/a ratio [Figure 3(b)] approaches the ideal value of single crystal ZnO (1.63) with increasing Si substitution. This trend again indicates that strain in the lattice is decreased. Strain was calculated from the XRD data using Williamson-Hall equation which confirms that strain decreases with increasing Si content [figure 4(b)].

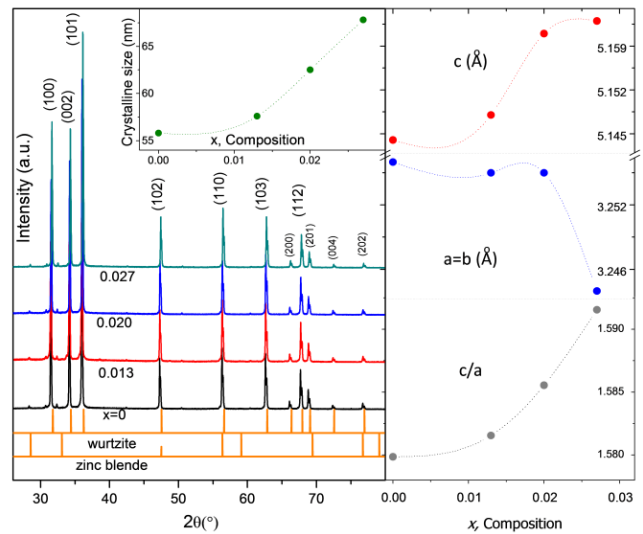

Figure 3: (a) XRD pattern of ZS0, ZS1, ZS2 & ZS3 (b) lattice parameter and c/a with x

A similar strain analysis was done using Raman analysis. Raman spectroscopy [figure 4(a)] revealed prominent $2E^2_{low}$, $E^2_{high} - E^2_{low}$ and $E^2_{high}$ phonon modes visible at ~201, 325 and 435 cm$^{-1}$. Two weak peaks at ~377 and 576 cm$^{-1}$ corresponding to $A1(TO)$ and $E1(LO)$ modes also appear in the Raman spectra. It has been already reported that the most intense $E^2_{high}$ mode is related to good crystallinity of wurtzite ZnO [21]. The intensity of the $E^2_{high}$ mode increases indicating that the crystalline quality increases.

$E^2_{low}$ are two non-polar modes associated with relative vibrations of the oxygen and the zinc sublattices in two different orientations. Note that the $E1(LO)$ mode which is related to oxygen/zinc vacancies is very weak indicating lesser contribution from such vacancies. In fact this also matches our results from photoluminescence studies discussed later. Strain was calculated from the A1 (TO) peak, which reflects the strength of polar lattice bonds [21], in x- direction and y-direction using formula [22]:

$$\epsilon_{xx} = \epsilon_{yy} = -\frac{\Delta\omega\,(A1(TO))}{b-a(\frac{C33}{C13})} \times \frac{C33}{2C13}$$

where $\Delta\omega=\omega-\omega_0$, deformation potential constant of $A_1(TO)$ mode i.e., a= -774 cm$^{-1}$ and b= -375 cm$^{-1}$, elastic stiffness constant of $C_{33}$ and $C_{13}$ are 216 and 104 GPa respectively. We found that strain decreases with increasing Si substitution in close agreement with our XRD results [figure 4(c)].

oxygen vacancies as revealed from our previous study [1]. To further confirm reduction in latice disorderness or strain, Urbach energy, which is a measure of structural disorderness, was estimated [figure 5 (c)]. The absorption coefficient at the photon energy below the optical gap (tail absorption) depends exponentially on the photon energy:

$\alpha(\hbar v)= \alpha_0 \exp(\hbar v/Eu)$

Where Eu is called Urbach energy and $\alpha_0$ is constant. Urbach energy decreases from 58 meV in ZS0 to 44 meV in ZS3, i.e. a reduction in Urbach tail width, ΔU, which effectively increases the bandgap [figure 5(b)]. Such reductions are associated with reduction in lattice irregularity and more crystalline material is achieved with silicon substitution. These results agree with strain analysis done from XRD and Raman studies.

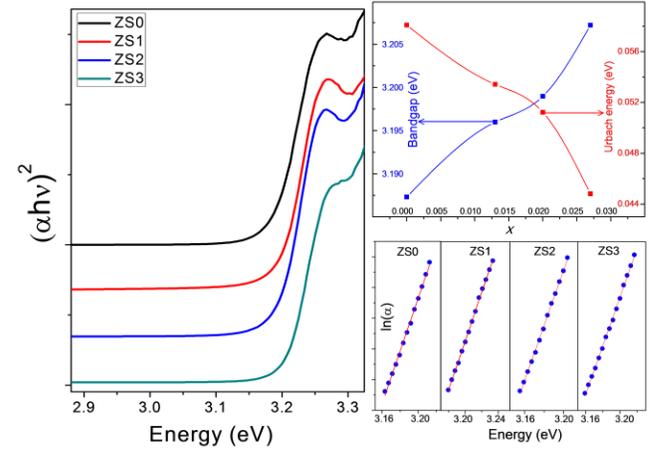

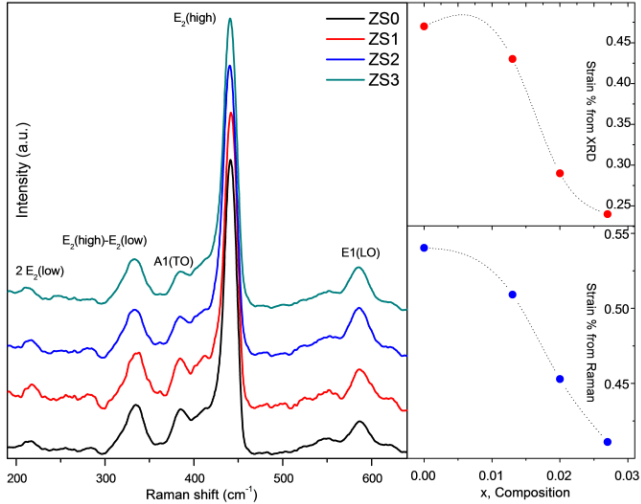

Figure 4: (a) Raman pattern of ZS0, ZS1, ZS2 & ZS3 (b) strain with x from XRD (c) strain with x from Raman

Optical transmittance shows increasing band gaps from 3.18eV, as in ZS0, steadily to 3.21eV [Figure 5a] as in ZS3 due to decrease in zinc and

Figure 5: (a) Optical transmittance of ZS0, ZS1, ZS2 & ZS3 (b) Bandgap and Urbach Energy with x (c) ln(α) with x

The strongest evidence of reduction of defects comes from the room temperature photoluminescence spectra [Figure 6 (a)]. The spectra consist of a near band UV emission (NBE) at ~390nm and a deep level visible (DLE) emission at approximately 500 nm. The DLE can be attributed to a radiative recombination of a photo generated hole with an electron occupying a surface oxygen vacancy. With increasing doping, the DLE intensities show a gradual decrease in luminescent intensity as shown in spectra. Extra charge of $Si^{4+}$ can retain more oxygen in the lattice thereby reducing oxygen vacancies defects in the system[1]. This is reflected in reduction of the DLE intensity. The bandgap becomes less populated by defect states thereby increasing the optical transparency to visible light with increasing $Si^{4+}$ substitution. However, the absorbance in the UV region remains unaltered. For quantitative analysis of defects states, photoluminescence spectra was fitted by multiple gaussian peaks.

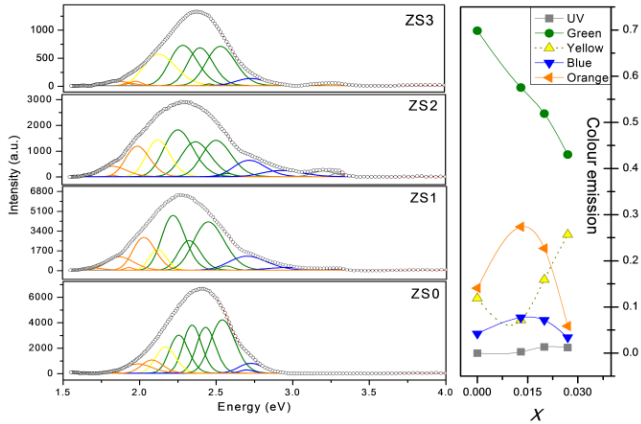

Figure 6: (a) Photoluminescence of Gaussian fitting of ZS0, ZS1, ZS2 & ZS3 (b) $P_{colour}$ with x

To understand the contribution of each component color in the total spectrum we calculated the color fraction, $P_{color}$, of each color given by $P_{color} = A_{color} / A_{total}$, where, $A_{total} = \Sigma A_{color}$, the total area under the spectrum. Green emission which is related to oxygen vacancies [figure 6(b)] reduced drastically. Other emission blue, yellow, orange related to zinc vacancies, interestitial defects have decreased nominally. Defects introduces strain in the lattice and one of the main reason for lattice distortion [23]. Reduction in oxygen vacancies defects in ZnO lattice due to extra charge present on silicon might have contributed towards decrease in strain and improvement in lattice irregularity.

Thus, a higher charged and lesser ionic radii $Si^{4+}$ incorporation in ZnO, decreases oxygen and zinc vacancy-related defects, providing strain relaxation to host lattice, thereby increasing the c/a ratio (approaching towards the ideal crystalline value ~1.63) as well as the particle size which is also reflected in the changes in lattice phonon modes.

The reduction in oxygen vacancies and thereby strain increases the DC conductivity of the material drastically by a factor of $\sim 10^4$. DC-conductivity and carrier concentration was estimated by studying current-voltage (I-V) characteristics and Hall measurements of the samples [figure 7(a, b, c)]. DC conductivity and carrier concentration, both increase with increase in Si incorporation. However, the DC conductivity remains nearly invariant for x=0.013, but thereafter increases exponentially. On the other hand from our Hall measurements even with x=0.013 substitution there is a sudden jump of carrier concentration. As conductivity and also carrier concentration are related to the mobility it is evident that the amount of Si content in the samples plays an important role on the mobility of the samples. At x=0.013 substitution the sudden jump of carrier concentration may be attributed to Si-doping, Si being in the $Si^{4+}$ state substituting a $Zn^{2+}$ ion, acts as a donor to the ZnO lattice. These donor sites provide extra electrons to the ZnO lattice. At the same time the extra charge also removes oxygen deficiency thereby bringing regularity in the lattice (c/a ratio approaching towards ideal crystalline values); thereby less scattering of electrons [24]. It seems for samples with x>0.013 the regularity is enhanced which improves the transport properties of the materials. Similar reports of enhanced conduction on account of increased carrier concentration due to extra charge of $Si^{4+}$ [10-11, 25-26] has been observed but a correlation with the reduction in strain in the lattice has not been made.

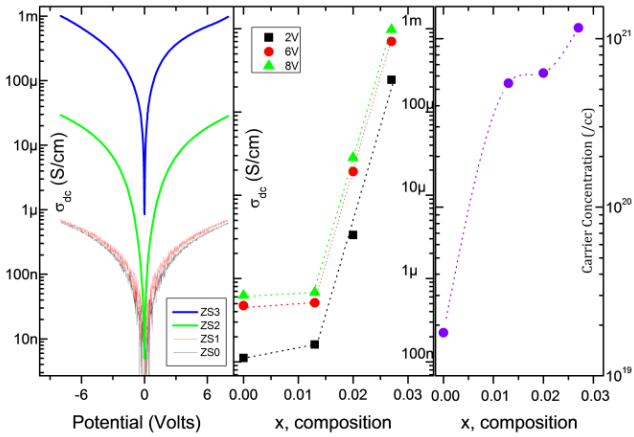

Figure 7: (a) Conductivity at voltage range of -8V to 8V (b) Conductivity with x (c) Carrier concentration changes with x.

Thus strain reduction in ZnO leads to improvement in sample crystallinity, enhancing conductivity and visible light transparency. Hence $Zn_{1-x}Si_xO$ is a UV-resistant optically transparent material. Good electrical conductivity and visible light transparency make this material fit for TCOs application.

Such an important material which is not only easily synthesized and also not expensive can be widely used as a conductive oxide which is also transparent. Thus we need to understand the materials' sensitivity to high humid conditions. Figure 8 (a, b, c, d) shows dynamic humidity response of pure and doped ZnO at room temperature. The resistance of all samples increases with humidity change from 45% to 75% and it is repeatable to next five cycles. Using the relation, $S = \Delta R/\Delta R_H$, the humidity sensitivity, S, was calculated. Sensitivity goes on decreasing with increasing silicon substitution mainly because the change of resistance is reducing with doping. The sensitivity was calculated to be have a decreasing trend 60 Ω/%RH, 46 Ω/%RH, 21 Ω/%RH and 1.67 Ω/%RH respectively (figure 8(e)) for ZS0, ZS1, ZS2 and ZS3.

In ZS0, the resistance increases from 29KOhms by 1.5KOhms (~5.2% increase). The same is about 4.5% for ZS1. For ZS2 and ZS3, it reduces further. The response time (time required to reach 90% of final value of resistance in the presence of humid air) and recovery time (time required to reduce the resistance to 10% of saturation value in the presence of dry air) was recorded and shown in fig 8(f).

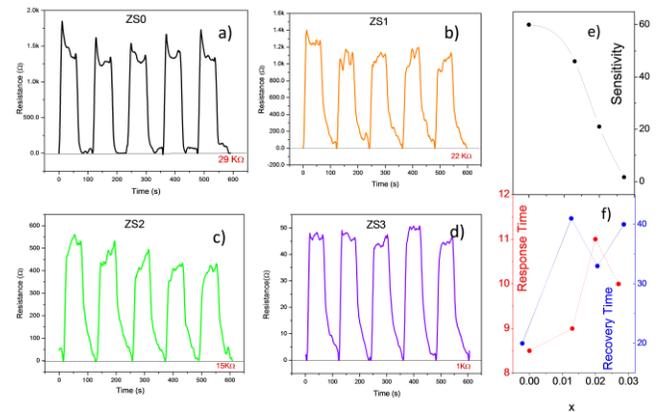

Figure 8: Dynamic response towards humidity of a)ZS0 b) ZS1 c) ZS2 d) ZS3 (e) sensitivity changes with x (f) response and recovery time with x.

The response time of ZS0, ZS1, ZS2 and ZS3 are 8.5s, 9s, 12s and 8s respectively whereas the recovery time is 20s, 41s, 32s and 40s respectively (Figure 8(f)). Response time is faster than recovery time in all cases but the recovery time becomes slower with doping which clearly signifies the decaying of sensing properties of ZnO. This value shows the reduction of moisture content in ZnO lattice due to silicon incorporation. Note that pure ZnO is itself highly sensitive to moisture and further inform that this sensitivity decreases with Si-doping. The sensing mechanism is based upon interaction between ZnO surface and water vapour. The hydrogen sites of water molecules are positively charged due to high electronegative nature of oxygen compared to hydrogen. These charged hydrogen sites capture free electrons from the conduction band of ZnO which reduces the current density of the system. Thus, the resistance of the system increases with increasing RH level. However with Si-incorporation the number of donor electrons will be more and hence the relative change in resistivity will be less. Many reports [27-34] claimed increase in RH level decreases the resistance which is explained by capillary condensation produced on ZnO surface. This capillary condensation process takes place for nanoparticles of 2 to 100nm whereas for ZnO nanorods (size>100 nm) inverse behavior is observed [27, 35-36]. Our results are similar to as reported for ZnO nanorods which may be due to the large particle size as revealed from SEM image.

The adsorption of water vapour on surface depends upon adsorption site which is directly proportional to number of pores present. Also oxygen vacancies near the surface of nanoparticles [37], behaves as adsorption sites. PL results suggests that the reduction of oxygen vacancies with Si doping. Si incorporation does not significantly affect the surface porosity as revealed from SEM studies. Thus, in our study, surface porosity may not be a reason behind lower sensitivity and moisture resistant property of Si doped ZnO TCOs.

**Conclusion:**

Proper doping of a higher charged and lesser ionic radii $Si^{4+}$ incorporation in ZnO, decreases oxygen and zinc vacancy-related defects, improving the optical transitivity in the visible region, providing strain relaxation to host lattice, thereby increasing the c/a ratio (approaching towards the ideal crystalline value ~1.63) as well as the particle size. This is also reflected in the changes in lattice phonon modes.

$Si^{4+}$ acts as a donor to the ZnO lattice, providing extra electrons to the ZnO lattice which enhances carrier concentration thereby increasing DC conductivity of the material drastically by a factor of ~$10^4$. The better optical transparency and high conductivity makes it a good TCO. With increasing silicon doping, the surface becomes less adsorbent of water molecules. Hence, conductivity remains less variant. The sensitivity towards water hence is drastically reduced making $Zn_{1-x}Si_xO$ an ideal candidate of moisture resistant probable TCO.

**Acknowledgement:**

We are thankful to SIC (IITI) for providing FESEM facility. The authors thank Dr. Vipul Singh for providing PL spectroscopy and Dr. Pankaj Sagdeo for providing UV–Vis facility.